\documentstyle[epsfig,prl,aps]{revtex}
\begin{document}
\draft
\twocolumn[\hsize\textwidth\columnwidth\hsize\csname
@twocolumnfalse\endcsname
\title{ Cosmological term as a source of mass}
\author{Irina Dymnikova
}
\address{Department of Mathematics and Computer Science,
         University of Warmia and Mazury,\\
Zolnierska 14, 10-561 Olsztyn, Poland; e-mail: irina@matman.uwm.edu.pl}

\maketitle

\begin{abstract}

In the spherically symmetric case the dominant energy condition,
together with the requirement of regularity at the center, asymptotic
flatness and finiteness of the ADM mass,
defines the family of asymptotically flat globally regular  solutions
to the Einstein equations which includes the class
of metrics  asymptotically de Sitter as $r\rightarrow 0$.
The source term corresponds to
an $r-$dependent cosmological term
$\Lambda_{\mu\nu}$  invariant under boosts in the radial direction
and evolving from the de Sitter vacuum $\Lambda g_{\mu\nu}$ in the origin
to the Minkowski vacuum at infinity. The ADM mass
is related to cosmological term by
$m=(2G)^{-1}\int_0^{\infty}{\Lambda_t^t r^2dr}$, with
de Sitter vacuum replacing a central singularity
at the scale of symmetry restoration.
Space-time symmetry changes smoothly from the de Sitter group near the center
to the Lorentz group at infinity through radial boosts in between.
In the range of masses $m\geq m_{crit}$, de Sitter-Schwarzschild geometry
describes  a vacuum nonsingular black hole ($\Lambda$BH),
and for $m<m_{crit}$ it describes G-lump - a vacuum selfgravitating
particlelike structure without horizons.
Quantum energy spectrum of G-lump is shifted down by the binding
energy, and zero-point vacuum mode is fixed at the value
corresponding (up to the coefficient)
to the Hawking temperature from the de Sitter horizon.

\end{abstract}


\vskip0.2in
]

{\bf Introduction -} In 1917 Einstein introduced a cosmological term
into his equations
describing gravity as spacetime geometry (G-field) generated by matter
$$
G_{\mu\nu}=-8\pi G T_{\mu\nu}\eqno(1)
$$
to make them consistent with Mach's principle which was
one of his primary motivations \cite{eins}.
Einstein's formulation of Mach's principle was that some matter has
the property of inertia only because there exists also some other matter
in the Universe (\cite{sciama}, Ch.2). When Einstein found that Minkowski
geometry is the regular solution to (1)
perfectly describing
inertial motion in the absence of any matter, he modified his equations
by adding the cosmological term $\Lambda g_{\mu\nu}$
in the hope that modified equations
$$
G_{\mu\nu}+\Lambda g_{\mu\nu}=-8\pi G T_{\mu\nu}\eqno(2)
$$
will have reasonable regular solutions only when matter is present
- if matter is the source
of inertia, then in case of its absence there should not be any inertia
\cite{bondi}. The primary task of $\Lambda$ was thus to eliminate inertia
in case when matter is absent by eliminating regular G-field solutions in case
when $T_{\mu\nu}=0$.

Soon after introducing $\Lambda g_{\mu\nu}$, in the same year 1917,
de Sitter found quite reasonable solution with $\Lambda g_{\mu\nu}$
and without $T_{\mu\nu}$, whose nowadays triumphs are well known.
The story of abandoning $\Lambda$ by Einstein is also widely known,
although typically with the accent on successes of FRW cosmology.
This somehow left in shadow the basic sense of his idea of introducing
$\Lambda$ as a quantity which has something in common with inertia.
The question - can it be possible to find some constructive way
connecting them? - seems to be related to the other
Einstein's profound proposal, suggested in 1950, to describe elementary
particle by regular solution of nonlinear field equations as "bunched
field" located in the confined region where field tension and energy
are particularly high \cite{bunch}. The possible way to such
a structure of gravitational origin whose mass is related to $\Lambda$
and whose regularity is related to this fact,
can be found in the Einstein field equations (1)  and in the Petrov
classification for $T_{\mu\nu}$ \cite{petrov}.

The aim of this paper is to show that in the spherically symmetric
case with the requirements of A) asymptotic flatness and finiteness
of the mass, B) regularity of metric and density at $r\rightarrow 0$,
and C) the dominant energy condition for a source term, there exists
the class of globally regular solutions, asymptotically
Schwarzschild at infinity, with de Sitter vacuum replacing a singularity.

\vskip0.1in
{\bf De Sitter-Schwarzschild geometry -}
A static spherically symmetric line element can be written in
the standard form (see, e.g., \cite{tolman}, p.239)
$$
ds^2 = e^{\mu(r)}dt^2 - e^{\nu(r)} dr^2 - r^2 d\Omega^2\eqno(3)
$$
where $d\Omega^2$ is the metric of a unit 2-sphere.

The Einstein equations (1) reduce
to (\cite{tolman}, p.244)
$$
8\pi G T_t^t = 8\pi G\rho(r)= e^{-\nu}\biggl(\frac{{\nu}^{\prime}}{r}
-\frac{1}{r^2}\biggr)
+\frac{1}{r^2}\eqno(4)$$
$$8\pi G T_r^r =-8\pi G p_r(r)= -e^{-\nu} \biggl(\frac{{\mu}^{\prime}}{r}
+\frac{1}{r^2}\biggr)
+\frac{1}{r^2}\eqno(5)$$
$$8\pi G T_{\theta}^{\theta}=8\pi G T_{\phi}^{\phi}=-8\pi G p_{\perp}(r)=$$
$$-e^{-\nu}\biggl(\frac{{{\mu}^{\prime\prime}}}{2}+\frac{{{\mu}^{\prime}}^2}{4}
+\frac{({{\mu}^{\prime}-{\nu}^{\prime}})}{2r}-\frac{{\mu}^{\prime}
{\nu}^{\prime}}{4}\biggr)\eqno(6)
$$
Here $\rho(r)=T^t_t$ is the energy density (we adopted $c=1$
for simplicity), $p_r(r)=-T^r_r$ is the radial pressure,
and $p_{\perp}(r)=-T_{\theta}^{\theta}=-T_{\phi}^{\phi}$
is the tangential pressure for a perfect fluid (\cite{tolman}, p.243).
A prime denotes differentiation
with respect to $r$.

To investigate this system in the case of different principal pressures
we impose requirement of regularity at the center of both density and metric,
and the energy dominant condition on the stress-energy tensor.

The dominant energy condition $T^{00}\geq|T^{ab}|$ for each $a,b=1,2,3$,
which holds if and only if \cite{HE}
$$\rho\geq0;~~~~-\rho\leq p_k\leq \rho;~~~~k=1,2,3\eqno(7)$$
implies that the local energy density is non-negative and each
principal pressure never exceeds the energy density.

Integration of Eq.(4) gives \cite{wald}
$$e^{-\nu(r)}=1-\frac{2GM(r)}{r};~~M(r)
=4\pi\int_0^r{\rho(x)x^2dx}\eqno(8)$$
which has for large $r$ the Schwarzschild asymptotics
$e^{-\nu}=1-{2Gm}/{r}$,
where $m$ is given by
$$
m=4\pi\int_0^{\infty}{\rho(r) r^2 dr}\eqno(9)
$$
In the limit $r\rightarrow\infty$ the condition of fineteness
of the mass (9) requires density profile
$\rho(r)$ to vanish at infinity quicker than $r^{-3}$.
The dominant energy condition $p_k\leq\rho$,
requires both radial and tangential pressures
to vanish as $r\rightarrow\infty$. Then $\mu^{\prime}=0$
and $\mu=$const at infinity, and we impose the standard
boundary condition $\mu\rightarrow 0$ as $r\rightarrow \infty$
to have asymptotic flatness needed to identify (9) as the ADM mass \cite{wald}.
As a result we get the Schwarzschild asymptotics at inifinity
$$T_{\mu\nu}=0;~~ds^2=\biggl(1-\frac{2Gm}{r}\biggr)-
\frac{dr^2}{\biggl(1-\frac{2Gm}{r}\biggr)}-r^2d\Omega^2\eqno(10)
$$
From Eqs.(4)-(6) we derive the equation (see also \cite{apj})
$$
p_{\perp}=p_r+\frac{r}{2}p_r^{\prime}+(\rho+p_r)\frac{G M(r)+4\pi G r^3 p_r}
{2(r-2G M(r))}\eqno(11)
$$
which is generalization of the Tolman-Oppenheimer-Volkoff
equation (\cite{wald}, p.127)
to the case of different principal pressures, and
the equation
$$ T_t^t-T_r^r=p_r+\rho=
\frac{1}{8\pi G}\frac{e^{-\nu}}{r}(\nu^{\prime}+\mu^{\prime})\eqno(12)
$$
From Eq.(8) we see that for any regular value of $e^{\nu(r)}$
we must have $M(r)=0$ at $r=0$\cite{oppi}, and that $\nu(r)\rightarrow 0$
as $r\rightarrow \infty$.
The dominant energy condition allows us to fix asymptotic behaviour
of a mass function and of a metric at approaching
the regular center.
Requirement of regularity of density $\rho(r=0)<\infty$,
leads, by the dominant energy condition $p_k\leq\rho$, to regularity
of pressures. Requirement of regularity of
the metric, $e^{\nu(r)}<\infty$, leads then,
by (12), to $\nu^{\prime}+\mu^{\prime}=0$
and $\nu+\mu=\mu(0)$ at $r=0$ with $\mu(0)$ playing the role
of the family parameter.

 The example of GR solution from this family is boson
stars \cite{boson} (for review \cite{Mielke1,Mielke})
which are  completely regular configurations without horizons
generated by self-gravitating massive complex scalar field
whose stress-energy tensor is essentially anisotropic,
$p_r\neq p_{\perp}$.

The weak energy condition,
$T_{\mu\nu}\xi^{\mu}\xi^{\nu}\geq 0$
for any timelike vector $\xi^{\mu}$, which
is satisfied
if and only if
$\rho\geq 0; \rho + p_k \geq 0, k=1,2,3$
and which is contained in the dominant energy condition\cite{HE},
defines, by Eq.(12), the sign of the
sum $\mu^{\prime}+\nu^{\prime}$. In the case when $e^{\nu (r)}>0$
everywhere, it demands  $\mu^{\prime}+\nu^{\prime}\geq 0$ everywhere.
In case when $e^{\nu (r)}$ changes sign, the function
$T_t^t-T_r^r$ is zero, by Eq.(12), at the horizons where $e^{-\nu}=0$.
In the regions inside the horizons, the radial coordinate $r$
is timelike and $T_t^t$ represents a tension,
$p_r=-T_t^t$, along the axes of the spacelike 3-cylinders of constant time
$r$=const \cite{werner}, then $T_t^t-T_r^r=-(p_r+\rho)$,
and the weak energy condition  still
demands $\nu^{\prime}+\mu^{\prime} \geq 0$ there.
As a result the function $\mu+\nu$ is a function growing from $\mu=\mu(0)$
at $r=0$ to $\mu=0$ at $r\rightarrow\infty$, which gives $\mu(0)\leq 0$.

This range of family parameter includes the value $\mu(0)=0$,
which corresponds to  $\nu+\mu=0$ at the center.
In this case the function $\phi(r)=\nu(r)+\mu(r)$ is zero at $r=0$ and
at $r\rightarrow\infty$, its derivative is non-negative,
it follows that $\phi(r)=0$, i.e., $\nu(r)=-\mu(r)$ everywhere.
The weak energy condition defines also equation of state and
thus asymptotic behaviour as $r\rightarrow 0$.
The function $\phi(r)=\mu(r)+\nu(r)$, which is equal zero
everywhere for $0\leq r<\infty$, cannot have extremum at $r=0$, therefore
$\mu^{\prime\prime}+\nu^{\prime\prime}=0$ at $r=0$ (this is easily
proved by contradiction using the Maclaurin rule
for even derivatives in the extremum).
It leads, by using L'Hopital rule in Eq.(12), to $p_r+\rho=0$
at $r=0$.
In the limit $r\rightarrow 0$
Eq.(11) becomes $p_{\perp}=-\rho-\frac{r}{2}\rho^{\prime}$.
The energy dominant condition (7) requires
$\rho^{\prime}\leq 0$, while regularity of $\rho$ requires
$p_k+\rho<\infty$ and thus $|\rho^{\prime}|<\infty$.
Then the equation of state near the center becomes
$p=-\rho$, which gives de Sitter asymptotics as $r\rightarrow 0$
$$ ds^2=\biggl(1-\frac{r^2}{r_0^2}\biggr)dt^2
-\frac{dr^2}{\biggl(1-\frac{r^2}{r_0^2}\biggr)}-r^2d\Omega^2\eqno(13)$$
$$
T_{\mu\nu}=\rho_0 g_{\mu\nu}; ~~~\rho_0=(8\pi G)^{-1}\Lambda;~~
r_0^2=\frac{3}{\Lambda}\eqno(14)
$$
where $\Lambda$ is the value of cosmological constant at $r=0$.

Summarizing, we conclude that if we require asymptotic flatness,
regularity of a density and metric  at the center and
finiteness of the ADM mass
$$e^\nu(r\rightarrow 0)<\infty;~~\rho(r\rightarrow 0)<\infty;~~
m<\infty\eqno(15)$$
then the dominant energy condition
defines the family of asymptotically flat solutions with the
regular center which includes the class of metrics
$$e^{\mu (r)}=e^{-\nu (r)}=g(r)=1-2G M(r)/r;~~~T_t^t=T_r^r\eqno(16)
$$
with $M(r)$ given by Eq.(8), whose behaviour in the
origin - asymptotically de Sitter as $r\rightarrow 0$, is defined
by  the weak energy condition. Note, that we need the dominant energy
condition $p_k\leq\rho$ only to restrict principal pressures by density
whose regularity is postulated. If we postulate regularity also for
pressures, then the weak energy condition is enough to distinguish
the class of metrics (16) asymptotically de Sitter in the origin,
as the member of family of asymptotically flat solutions with
the regular center.

For this class the equation of state is
$$p_r=-\rho:~~~p_{\perp}=-\rho-(r/2)\rho^{\prime}\eqno(17)$$

The source term connects de Sitter vacuum $T_{\mu\nu}=\rho_0 g_{\mu\nu}$
in the origin with the Minkowski vacuum $T_{\mu\nu}=0$ at infinity,
and generates de Sitter-Schwarzschild geometry \cite{me96}
asymptotically de Sitter as $r\rightarrow 0$ and asymptotically
Schwarzschild as $r\rightarrow\infty$.

The weak energy condition $p_{\perp}+\rho\geq 0$
gives $\rho^{\prime}\leq 0$,
so that it demands monotonic
decreasing of a density profile.
By Eq.(6) it leads
to the important fact that,
except the point $r=0$ where $g(r)$ has the maximum,
in any other extremum $g^{\prime\prime}> 0$, so that the
function $g(r)$ has in the region $0<r<\infty$ only minimum
and the metric (16) can have not more than two horizons.

Indeed, for the metric (16) the Eq.(6) reduces to
$$8\pi Gp_{\perp}=-\frac{GM^{\prime\prime}}{r}\eqno(18)$$
The derivative of the mass function $M(r)$ is always positive
since the density is positive
($M^{\prime}=4\pi \rho r^2$); the function $M^{\prime}(r)$
can have only maximum and only one at the point $r_c$ where
$p_{\perp}(r_c)=0$ and hence $M^{\prime\prime}(r_c)=0$
(by Eq.(17) tangential pressure
$p_{\perp}(r)$ can change sign only once.) At the extremum
$r=r_m$ of the metric function $g(r)$, Eq.(18) takes the form
$8\pi G p_{\perp}(r_m)=\frac{g^{\prime\prime}(r_m)}{2}$.
In the region $0\leq r< r_c$, the derivative $M^{\prime\prime}>0$,
and hence for extrema $r_m$ in this range
$g^{\prime\prime}(r_m)<0$, i.e. in this region there exists only
maximum (and only one, this is the maximum at $r=0$).
In the region $r_c< r\leq\infty$, the second mass derivativeis negative,
$M^{\prime\prime}<0$, and for extrema from this region
$g^{\prime\prime}(r_m)>0$, i.e. metric function $g(r)$ can have
here only minimum (and only one).

To find explicit form of $M(r)$ we have to choose the density profile
leading to the needed behaviour of $M(r)$ as $r\rightarrow 0$,
$M(r)\simeq{(4\pi r^3 /3) \rho_0}$. The simplest choice \cite{me92}
$$
\rho(r)=\rho_0 e^{-r^3/r_0^2 r_g}
=\rho_0 e^{-\frac{4\pi\rho_0}{3m}r^3}\eqno(19)
$$
can be interpreted \cite{me96} as due to vacuum polarization
in the spherically symmetric gravitational field as described semiclassically
by the Schwinger formula $w\sim{e^{-F_{crit}/F}}$ (see, e.g., \cite{igor})
with tidal forces $F\sim{r_g/r^3}$ and $F_{crit}\sim{1/r_0^2}$, in agreement
with the basic idea suggested by Poisson and Israel that
in Schwarzschild-de Sitter transition spacetime geometry can be
self-regulatory and describable semicalssically down to a few Planckian radii
by the Eisntein equations with a source term representing vacuum polarization
effects \cite{werner}. This density profile gives
$$
M(r)=m(1-e^{-r^3/r_0^2 r_g})\eqno(20)
$$
The key point is existence of two horizons, a black hole event horizon
$r_{+}$ and an internal horizon $r_{-}$. A critical value
of a mass parameter exists, $m_{crit}$,
at which the horizons come together and which puts a lower limit
on a black hole mass \cite{me96}.
 For the model (19)
$$
m_{crit}\simeq{0.3 m_{Pl}\sqrt{\rho_{Pl}/\rho_0}}\eqno(21)
$$
De Sitter-Schwarzschild configurations are shown in Fig.1.
\begin{figure}
\vspace{-8.0mm}
\begin{center}
\epsfig{file=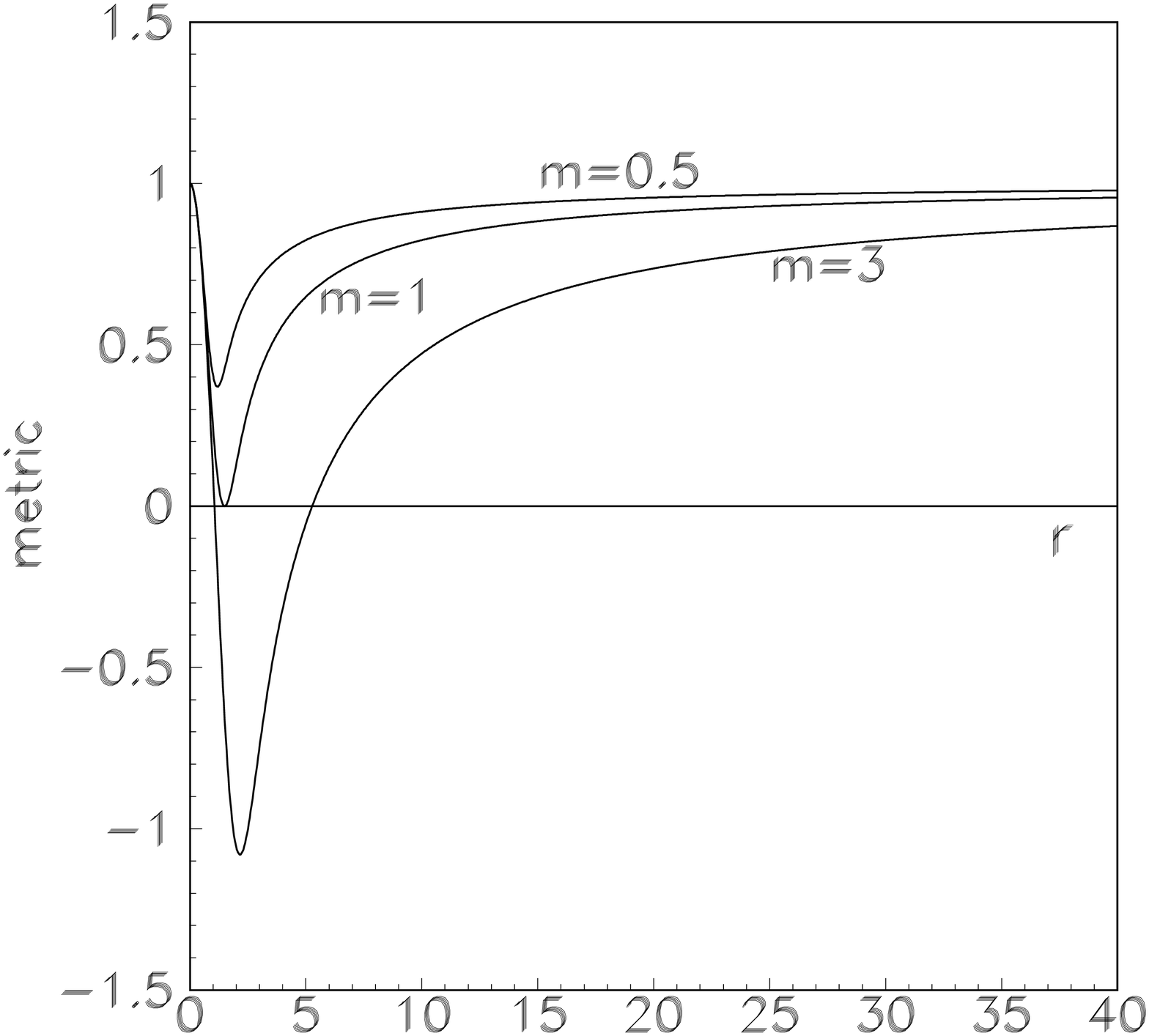,width=8.0cm,height=5.5cm}
\end{center}
\caption{
The metric $g(r)$ for de Sitter-Schwarzschild
configurations plotted for the case of the density profile (19).
The mass $m$ is normalized to $m_{crit}$.
}
\label{fig.1}
\end{figure}

This class of metrics is easily extended to the case of nonzero
background cosmological constant $\lambda$, by introducing
$T_t^t(r)=\rho(r)+(8\pi G)^{-1}\lambda$. Then the metric function $g(r)$
in Eq.(16) is given by
$g(r)=1-\frac{2GM(r)}{r} -\frac{\lambda r^2}{3}$\cite{us97}.
Stress-energy tensors for this class of metrics
$$
T_t^t=T_r^r;~~T_{\theta}^{\theta}=T_{\phi}^{\phi}\eqno(23)
$$
belong to the Petrov type [(II)(II)] \cite{petrov}.
 The first symbol in the brackets denotes the eigenvalue
related to the timelike eigenvector representing a velocity.
Parentheses combine equal eigenvalues.
A comoving reference frame is defined uniquely
 if and only if none of the
spacelike eigenvalues coincides with
a timelike eigenvalue.

Stress-energy tensor (23) has an infinite set
of comoving reference frames, since it is
invariant under boosts in the radial direction, and
can be thus identified as describing a spherically symmetric vacuum
(an observer moving through such  a medium
cannot in principle measure the radial component of his velocity
with respect to it), i.e., vacuum
 with variable energy densitity and pressures, macroscopically
defined by the algebraic structure (23)
of its stress-energy tensor $T_{\mu\nu}^{vac}$ \cite{me92}.
 In the case of nonzero background $\lambda$
it connects in a smooth way two de Sitter vacua
with different values of cosmological constant.
This makes it possible to interpret $T_{\mu\nu}^{vac}$ as
corresponding to the extension of the algebraic structure
of the cosmological term from $\Lambda g_{\mu\nu}$ (with $\Lambda$=const)
to an $r$-dependent  cosmological term
$\Lambda_{\mu\nu}=8\pi G T_{\mu\nu}^{vac}$,
evolving from $\Lambda_{\mu\nu}=\Lambda g_{\mu\nu}$
as $r\rightarrow 0$ to $\Lambda_{\mu\nu}=\lambda g_{\mu\nu}$
as $r\rightarrow\infty$, and
satisfying the equation of state (17)
with $8\pi G \rho^{\Lambda}=\Lambda^t_t$,
$8\pi G p_r^{\Lambda}=-\Lambda^r_r$ and $8\pi G p_{\perp}^{\Lambda}
=-\Lambda^{\theta}_{\theta}=-\Lambda^{\phi}_{\phi}$ \cite{me00}.

In this paper we concentrate on de Sitter-Schwarzschild
geometry (16)
generated by a cosmological term $\Lambda_{\mu\nu}$ evolving
from the de Sitter vacuum $\Lambda_{\mu\nu}=\Lambda g_{\mu\nu}$
at $r=0$ to the Minkowski vacuum $\Lambda_{\mu\nu}=0$ at infinity.

For $m\geq m_{crit}$ de Sitter-Schwarzschild geometry describes the
vacuum nonsingular black hole ($\Lambda$BH) \cite{me92},
and global structure of spacetime, shown in Fig.2 \cite{me96},
contains an infinite
sequence of black and white holes
whose future and past singularities are replaced with regular cores
$\cal{RC}$
asymptotically de Sitter as $r\rightarrow 0$.

\begin{figure}
\vspace{-8.0mm}
\begin{center}
\epsfig{file=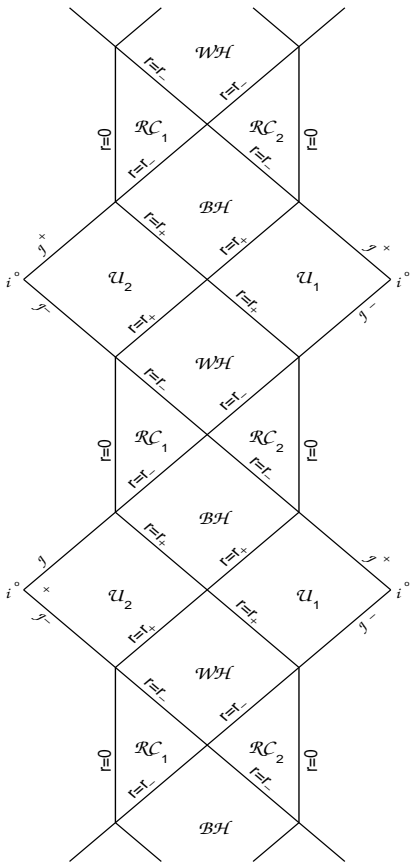,width=8.0cm,height=12.9cm}
\end{center}
\caption{Penrose-Carter diagram for $\Lambda$ black hole.}
\label{fig.2}
\end{figure}

A $\Lambda$BH emits Hawking radiation from both horizons (see Fig.3),
and configuration evolves towards a self-gravitating particlelike structure
without horizons (see Fig.1).
While a $\Lambda$BH loses mass, horizons come together and temperature
drops to zero \cite{me96}.
The Schwarzschild asymptotics
requires $T_{+}\sim{m^{-1}}\rightarrow 0$
as $m\rightarrow\infty$.
The temperature $T_{+}$ on BH horizon $r_{+}$
is positive by general
laws of BH thermodynamics \cite{wald}.
As a result the temperature-mass diagram has a maximum
between $m_{crit}$ and $m\rightarrow\infty$ \cite{me96}.
In a maximum a specific heat is broken and changes
sign testifying to a second-order
phase transition in the course of Hawking evaporation
(and suggesting symmetry restoration in the origin \cite{me97}).

\begin{figure}
\vspace{-8.0mm}
\begin{center}
\epsfig{file=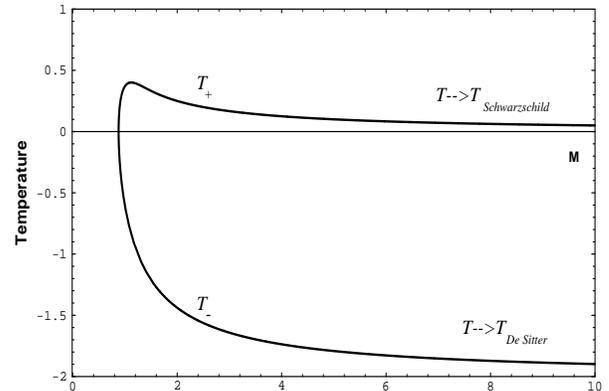,width=8.0cm,height=5.5cm,clip=}
\end{center}
\caption{
Temperature-mass diagram for $\Lambda$ black hole.}
\label{fig.3}
\end{figure}

For masses $m<m_{crit}$ de Sitter-Schwarzschild geometry describes
a self-gravitating particlelike vacuum structure without horizons,
globally regular and globally neutral.
 It resembles
Coleman's lumps - non-singular, non-dissipative solutions
of finite energy, holding themselves together by their
own self-interaction \cite{lump}.
Our lump is regular solution
to the Einstein equations, perfectly localized (see Fig.4) in a
region where field tension and energy are particularly high
(this is the region of the former singularity), so we can call it
G-lump.
\begin{figure}
\vspace{-8.0mm}
\begin{center}
\epsfig{file=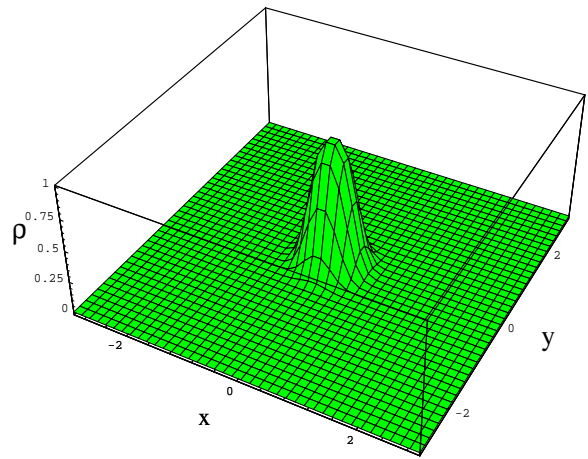,width=8.0cm,height=6.5cm}
\end{center}
\caption{
G-lump in the case $r_g=0.1r_0$ ($m\simeq{0.06 m_{crit}})$.
}
\label{fig.4}
\end{figure}

 It holds itself together by gravity
due to balance between gravitational attraction outside and
gravitational repulsion
inside of zero-gravity surface $r=r_c$ beyond which
the strong energy condition
of singularities theorems \cite{HE},
$(T_{\mu\nu}-T g_{\mu\nu}/2)\xi^{\mu}\xi^{\nu})\geq 0$,
is violated \cite{me96}.
The surface of zero gravity is defined by $2\rho+r\rho^{\prime}=0$.
It is depicted in Fig.5 together
with horizons and with the surface $r=r_s$ of zero scalar
curvature $R(r_s)=0$ which in the case of the density profile (19)
is given by
$$r_s=\biggl(\frac{4}{3}r_0^2 r_g\biggr)^{1/3}
=\biggl(\frac{m}{\pi\rho_0}\biggr)^{1/3}\eqno(24)$$
and confines about 3/4 of the mass $m$.
\begin{figure}
\vspace{-8.0mm}
\begin{center}
\epsfig{file=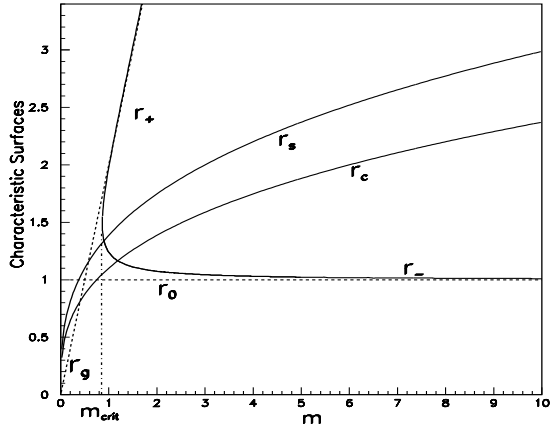,width=8.0cm,height=6.5cm}
\end{center}
\caption{
Horizons of $\Lambda$BH, surface of zero scalar curvature $r=r_s$
and surface of
zero gravity $r=r_c$.
}
\label{fig.5}
\end{figure}
\vskip0.1in
The mass of both G-lump and $\Lambda$BH is directly connected
to cosmological term $\Lambda_{\mu\nu}$ by the ADM formula (9)
which in this case reads
$$
m=(2 G)^{-1} \int_0^{\infty}{\Lambda_t^t(r) r^2 dr}\eqno(25)
$$
and relates mass to the de Sitter vacuum at the origin.

The Minkowski geometry allows existence of inertial mass
as the Lorentz invariant
$m^2=p_{\mu}p^{\mu}$  of a  test body.
High symmetry of this geometry allows both
existence of inertial frames and of quantity $m$ as the measure of inertia,
but geometry tells nothing about this quantity.

In the Schwarzschild geometry the parameter $m$ is
responsible for geometry, it is identified
as a gravitational mass of a source by asymptotic behavior
of the metric at infinity (see, e.g., \cite{wald}, p.124).
By the equivalence principle, gravitational mass is equal
to inertial mass (see, e.g., \cite{sciama}, Ch.4).
The inertial mass is represented thus by a purely geometrical quantity,
the Schwazschild radius $r_g$ which is just geometrical fact \cite{wheeler2},
but which does not tell yet anything about origin of a mass.

In de Sitter-Schwarzschild geometry the parameter $m$ is identified as
a mass by Schwarzschild asymptotics at infinity.
The geometrical fact of
this geometry
is that a mass is related to cosmological term, since Schwarzschild
singularity is replaced with a de Sitter vacuum.
The operation of introducing mass by the ADM formula (9) is impossible
in the de Sitter geometry.
 The reason is that symmetry of the source term
$T_{\mu\nu}=\rho_0 g_{\mu\nu}=(8\pi G)^{-1}\Lambda g_{\mu\nu}$ is too high.
It implies $\rho_{vac}$=const by virtue of the Bianchi identities
$G^{\mu\nu}_{;\nu}=0$.
In the case of geometry generated by the cosmological term $\Lambda_{\mu\nu}$,
symmetry of a source term is reduced from the full Lorentz group to
the Lorentz boosts in the radial direction only.
Together with asymptotic flatness this allows introducing a distinguished point
as the center of an object whose ADM mass is defined by the formula (25).
The reduced symmetry of a source and the
asymptotic flatness of geometry are responsible
for mass of an object given by (25).

This picture seems to be in remarkable conformity with the basic idea
of the Higgs mechanism for generation of mass via spontaneous
breaking of symmetry of a scalar field vacuum from a
false vacuum (where $T_{\mu\nu}=V(0)g_{\mu\nu}$, and $p=-\rho$),
to a true vacuum $T_{\mu\nu}=0$. In both cases de Sitter vacuum is involved
and vacuum symmetry is broken.
Even graphically the gravitational potential $g(r)$ resembles
a Higgs potential (see Fig.6).
\begin{figure}
\vspace{-8.0mm}
\begin{center}
\epsfig{file=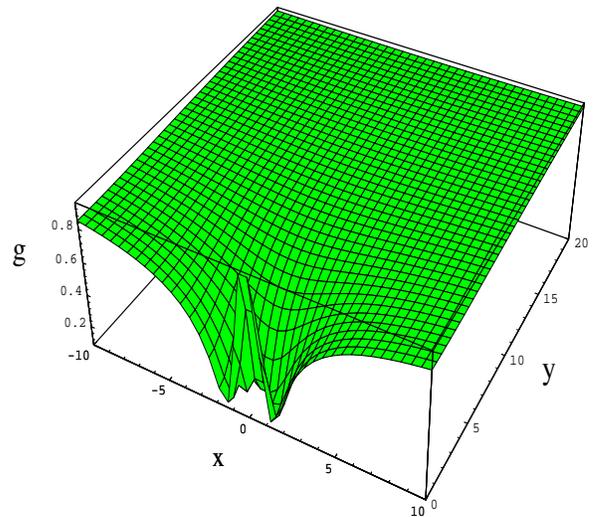,width=8.0cm,height=7.5cm}
\end{center}
\caption{The gravitational potential $g(r)$ for the case of
G-lump with the mass a little bit less than $m_{crit}$.
}
\label{fig.6}
\end{figure}

The difference is that in case of a mass coming from
$\Lambda_{\mu\nu}$ by (25), the gravitational potential $g(r)$
is generic, and
de Sitter vacuum supplies
a particle with mass via smooth breaking of space-time
symmetry from the de Sitter group in its center
to the Lorentz group at its infinity.

 This
leads to the natural assumption \cite{ethz} that whatever would be
particular mechanism for mass generation, a fundamental particle
(which does not display substructure, like a lepton or quark) may have
an internal vacuum core (at the scale where it gets mass) related
to its mass and a  geometrical size defined by gravity.
Such a core with de Sitter vacuum at the origin and Minkowski
vacuum at infinity can be approximated by de Sitter-Schwarzschild
geometry. Characteristic size in
this geometry is given by (24). It depends on vacuum density
at $r=0$ and presents modification of the Schwarzschild radius $r_g$
to the case when singularity is replaced by de Sitter vacuum.
While application of the Schwarzschild radius to elementary particle
size is highly speculative since obtained estimates are many orders
of magnitude less than $l_{Pl}$, the characteristic size $r_s$ gives
reasonable numbers (e.g., $r_s\sim{10^{-18}}$ cm for the electron getting
its mass from the vacuum at the electroweak scale)
close to estimates obtained in experiments (see Fig.7\cite{ethz}
where they are compared with electromagnetic (EM) and electroweak (EW)
experimental limits).
\begin{figure}
\vspace{-8.0mm}
\begin{center}
\epsfig{file=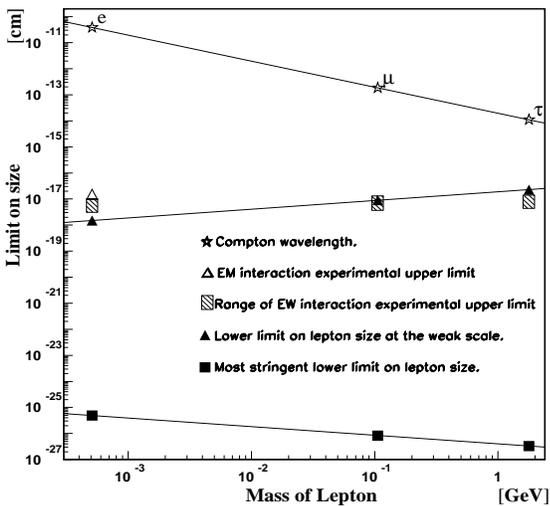,width=8.0cm,height=7.5cm}
\end{center}
\caption{Characteristic sizes for leptons \protect\cite{ethz}.
}
\label{fig.7}
\end{figure}

\vskip0.1in

The question always discussed in connection with cosmological
constant is zero-point vacuum energy.
Let us evaluate zero-point energy for G-lump,
which clearly represents an elementary spherically symmetric
excitation of a vacuum defined macroscopically by (23).

Since de Sitter vacuum is trapped within a G-lump,
we can model it by a spherical bubble
whose density decreases with a distance.
In the Finkelstein coordinates, its geometry is
described by the metric
$$
 ds^2=d\tau^2-\frac{2GM(r(R,\tau))}{r(R,\tau)}-r^2(R,\tau)d\Omega^2
\eqno(26)$$
The equation of motion
$\dot{r}^2 + 2r\ddot{r}-8\pi G\rho(r)r^2=f(R)$ \cite{us00},
where dot denotes differentiation with respect to $\tau$ and
$f(R)$ is constant of integration, has the first integral
$$
\dot{r}^2 - \frac{2GM(r)}{r}=f(R)\eqno(27)
$$
which resembles the equation of a particle in the potential
$V(r)=-\frac{GM(r)}{r}$, with the constant
of integration $f(R)$ playing the role of the total energy $f=2E$.
The Hamiltonian of this model is ${\cal{H}}
=\frac{\dot{r}^2}{2}-\frac{GM(r)}{r}$, the Lagrangian
${\cal{L}}=\frac{\dot{r}^2}{2}+\frac{GM(r)}{r}$, and the conjugate
momentum $p=\frac{\partial {\cal{L}}}{\partial {\dot{r}}}=\dot{r}$.

A spherical bubble can be described by the minisuperspace model
with a single degree of freedom \cite{vil}. The momentum operator
is introduced by $\hat{p}=-i{l_{Pl}}^2 d/dr$, and the equation
(27) transforms into the Wheeler-DeWitt equation in the
minisuperspace \cite{vil} which reduces to the Schr\"odinger
equation
$$
\frac{\hbar^2}{2m_{Pl}}\frac{d^2\psi}{dr^2}-(V(r)-E)\psi=0\eqno(28)
$$
with the potential (in the Planckian units)
$$
V(r)=-\frac{GM(r)}{r}\eqno(29)
$$
depicted in Fig.8.
\begin{figure}
\vspace{-8.0mm}
\begin{center}
\epsfig{file=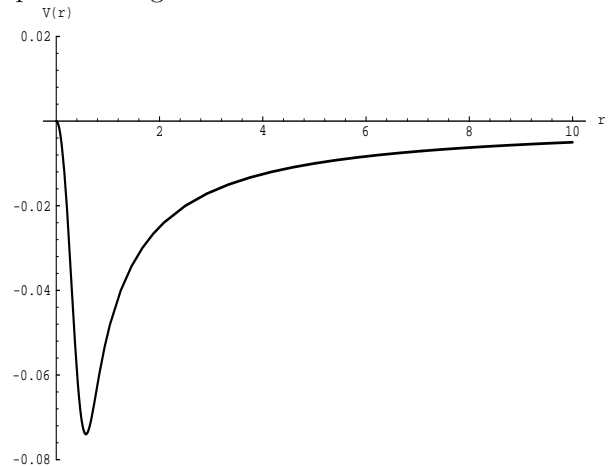,width=8.0cm,height=6.5cm}
\end{center}
\caption{The plot of the potential (29) for G-lump
with $r_g=0.1 r_0 (m\simeq 0.07 m_{crit})$.
}
\label{fig.8}
\end{figure}

Near the minimum $r=r_m$ the potential takes the form
$V(r)=V(r_m)+4\pi G p_{\perp}(r_m)(r-r_m)^2$.
Introducing the variable $x=r-r_m$
we reduce  Eq.(28) to
the equation for a harmonic oscillator
$$
\frac{d^2\psi}{dx^2}-\frac{m_{Pl}^2\omega^2 x^2}{\hbar^2}\psi
+\frac{2m_{Pl}\tilde{E}}{\hbar^2}\psi=0\eqno(30)
$$
where $\tilde{E}=E-V(r_m)$,
$\omega^2=\Lambda c^2 \tilde{p}_{\perp}(r_m)$, and
$\tilde{p}_{\perp}$ is the dimensionless pressure normalized to vacuum density
$\rho_0$ at $r=0$; for the density profile (19)
$\tilde{p}_{\perp}(r_m)\simeq{0.2}$.
The energy spectrum
$$
E_n=\hbar \omega \biggl(n+\frac{1}{2}\biggr)
-\frac{GM(r_m)}{r_m}E_{Pl}\eqno(31)
$$
is shifted down by the minimum of the potential $V(r_m)$ which represents
the binding energy.
The energy of zero-point vacuum mode
$$
\tilde{E}_0=\frac{\sqrt{3\tilde{p}_{\perp}}}{2}\frac{\hbar c}{r_0}\eqno(32)
$$
never exceeds the binding energy $V(r_m)$. It
remarkably agrees with the Hawking temperature from the de Sitter horizon
$kT_H=\frac{1}{2\pi}\frac{\hbar c}{r_0}$ \cite{GH}, representing the energy of
virtual particles which could become real in the presence of the horizon.
In the case of G-lump which is structure without horizons, kind of
gravitational vacuum exciton,
they are confined by the binding energy $V(r_m)$.
\vskip0.1in

The question of stability of de Sitter-Shwarzschild configurations
is currently under investigation. De Sitter-Schwarzschild black hole
configuration obtained by direct matching of the Schwarzschild
metric outside to de Sitter metric inside of a spacelike three-cylindrical
short transitional layer \cite{valera} is a stable configuration in a sense
that the three-cylinder does not tend to shrink down under
perurbations \cite{Poisson}.
De Sitter-Schwarzscild configurations considered above represent
general case of a smooth transition with a distributed density profile.
The heuristic argument in favour of their stability comes from comparison
of the ADM mass with the proper mass \cite{wald}
$\mu=4\pi\int_0^{\infty}{\rho(r)\biggl(1
-\frac{2 G M(r)}{r}\biggr)^{-1/2}r^2 dr}$
In the spherically symmetric case the ADM mass represents
the total energy, $m=\mu+$binding energy \cite{wald}.
In de Sitter-Schwarzschild geometry $\mu$ is bigger than $M$.
This suggests that the configuration might be stable
since energy is needed to break it up \cite{me00}.
Analysis of stability of a $\Lambda$BH as an isolated system by Poincare's
method, with the total energy $m$ as a thermodynamical variable
and the inverse temperature as the  conjugate variable \cite{Kaburaki},
shows immediately its stability with respect to spherically symmetric
perturbations. The analysis by Chandrasekhar method \cite{chandra}
is straightforward for a $\Lambda$BH stability to external perturbations,
in close similarity with the Schwarzschild
and Reissner-Nordstr\"om cases. The potential barriers
in one-dimensional wave equations
governing perturbations, external to the event
horizon, are real and positive, and stability follows from this fact
\cite{chandra}. Preliminary results suggest stability
also for the case of G-lump. In the context of
catastrophe-theory analysis, de Sitter-Schwarschild configuration resembles
high-entropy neutral type in the Maeda classification, in which a non-Abelian
structure may be approximated as a uniform vacuum density $\rho_{vac}$ within
a sphere whose radius is the Compton wavelength of a massive non-Abelian field,
and self-gravitating particle approaches the particle solution in the Minkowski
space \cite{Maeda}.
\vskip0.1in
{\bf Discussion -}The main result of this paper is that
there exists the class of globally regular solutions to the
minimally coupled GR equations (4)-(6),
with the algebraic structure of a source term (23),
interpreted as spherically symmetric vacuum with variable
density and pressure $T_{\mu\nu}^{vac}$ associated with a variable
cosmological term $\Lambda_{\mu\nu} =8\pi GT_{\mu\nu}^{vac}$,
whose asymptotics in the origin, dictated by the weak energy condition,
is the Einstein cosmological term $\Lambda g_{\mu\nu}$.
For this class the mass defined by
the standard ADM formula (9) is related (generically, since matter
source can be any from the considered class) to both de Sitter vacuum
trapped in the origin and to breaking of space-time symmetry.

In the inflationary cosmology which is based
on generic properties of de Sitter vacuum $\Lambda g_{\mu\nu}$
independently on where
$\Lambda$ comes from \cite{us}, several
mechanisms are investigated relating  $\Lambda g_{\mu\nu}$
to matter sources (for review see \cite{olive}).
Most frequently considered is a scalar field
$$S=\int{d^4 x \sqrt{-g}\biggl[R+(\partial \phi)^2-2V(\phi)\biggr]}\eqno(33)$$
where $R$ is the scalar curvature,
$(\partial \phi)^2=g^{\mu\nu}{\partial }_{\mu}\phi {\partial }_{\nu}\phi$,
with various forms for a scalar field potential  $V(\phi)$.

The question whether a regular black hole  can be obtained
as a false vacuum configuration described by (33),
has been addressed in the paper  \cite{dima},
where "the no-go theorem" has been proved:
Asymptotically flat regular black hole solutions are absent  in the
theory  (33) with any non-negative potential $V(\phi)$. This result
has been extended to the case of any $V(\phi)$ and any asymptotics
and then generalized to the case of a theory with the action
$S=\int{d^4 x\sqrt{-g}\biggl[R+F[(\partial \phi)^2,\phi]\biggr]}$,
where $F$ is an arbitrary function \cite{kirill},
to the multi-scalar theories of sigma-model type, and to scalar-tensor
and  curvature-nonlinear gravity theories \cite{kirill2}.
It has been shown that
the only possible regular solutions are either de Sitter-like
with a single cosmological horizon  or those
without horizons, including asymptotically flat ones. The latter
do not exist for $V(\phi)\geq 0$, so that
the set of causal false vacuum structures is the same as known for
$\phi=const$ case, namely Minkowski (or anti-de Sitter), Schwarzschild,
de Sitter, and Schwarzschild-de Sitter \cite{kirill,kirill2}, and thus does
not include de Sitter-Schwarzschild configurations.
However, as it was noted \cite{Schunck} three years before
formulating "no-go theorems", in the case of {\it complex} massive
scalar field the regular structures can be obtained in the minimally
coupled theory with positive $V(\phi)$ \cite{Schunck}. The best
example is boson stars (\cite{Mielke1,Mielke} and references therein),
but in this case algebraic structure of the stress-energy tensor
\cite{Mielke} does not satisfy Eq.(23), and asymptotics  at $r=0$
is not de Sitter.

\vskip0.1in
In a recent paper on $\Lambda$-variability,
Overduin and Cooperstock distinguished two basic approaches
to $\Lambda g_{\mu\nu}$ existing in the literature\cite{overduin}.
In the first approach
$\Lambda g_{\mu\nu}$ is shifted onto the right-hand side of the field
Einstein equations (2) and treated as a dynamical part of the matter content.
This approach, characterized by Overduin and Cooperstock as
being connected to dialectic materialism of the Soviet physics school,
goes back to Gliner who interpreted $\Lambda g_{\mu\nu}$
as vacuum stress-energy tensor \cite{EB},
to Zel'dovich who connected $\Lambda$ with the gravitational
interaction of virtual particles inhabitating vacuum \cite{zeld},
and to Linde who suggested that $\Lambda$
can vary in principle \cite{andrej}.
In contrast, idealistic approach prefers
to keep $\Lambda$ on the left-hand side of Eqs.(2) as geometrical
entity and treat it as a constant of nature \cite{phil}.

This classification suggests that any variable $\Lambda$ must
be identified with a matter, in such a case the best fit for
$T_{\mu\nu}^{vac}=(8\pi G)^{-1} \Lambda_{\mu\nu}$ would be
gravitational vacuum polarization in the spirit of Zel'dovich's idea
\cite{zeld} and Poisson and Israel self-regulatory picture \cite{werner}.
On the other hand nothing prevents from shifting $T_{\mu\nu}^{vac}$
from the right-hand to the left-hand side of Eqs.(1)
and treating $\Lambda_{\mu\nu}=8\pi G T_{\mu\nu}^{vac}$
as evolving geometrical entity.
(The Einstein field equations (1)
can be written in the four-indices form
$G_{\alpha\beta\gamma\delta}=-8\pi G T_{\alpha\beta\gamma\delta}$ as
the equivalence relations which put the matter and geometry in direct
algebraic correspondence \cite{us83}).

The considered connection between r-dependent cosmological term
$\Lambda_{\mu\nu}$ and the ADM mass seems to satisfy Einstein's
version of Mach's principle - no matter,
no inertia - if we explicitly separate two aspects of the problem
of inertia: existence of inertial frames and existence of inertial
mass. In empty space, $T_{\mu\nu}=0,\Lambda_{\mu\nu}=0$,
inertial frames exist due to high symmetry of Minkowski geometry,
but to prove it we need a measure of inertia, a test particle
with the inertial mass, i.e. a region in space where Minkowski
vacuum is a little bit disturbed. When a mass comes from cosmological
term, it is disturbed by $\Lambda_{\mu\nu}\neq 0$.
In other words, full symmetry of Minkowski spacetime is responsible
for existence of inertial frames, while its breaking to Lorentz boosts
in the radial direction only is responsible for inertial mass.

\vskip0.1in
{\bf Acknowledgement}

I am very grateful to Werner Israel for helpful discussions.
This work was supported by the Polish Committee for Scientific Research
through the Grant 5P03D.007.20, and through the grant for UWM.



\begin{references}
\bibitem{phil} It looks that the question of
where to put $\Lambda g_{\mu\nu}$ is kind of philosophical question
of what is primary. If we remind that dialectic materialism
is nothing but application of Hegel's dialectic idealism to matter,
then one is tempted to approach $\Lambda g_{\mu\nu}$  by the Hegel's laws
of a new triad, as a new quality ether appearing
in the new turn of cognitive helix, a Lorentz-invariant ether
with respect to which one cannot measure velocity in principle.
\end{references}
\end{document}